\documentstyle[aps]{revtex}
\input{epsf}

\def\IR{\relax{\rm I\kern-.18em R}}
\def\IN{\relax{\rm I\kern-.18em N}}
\def\cod{{\rm cod}}

\begin{document}
\draft
\title{Analysis of a three-component model phase diagram 
by Catastrophe Theory: Potentials with two Order Parameters.} 
%
%
\author{J. Gaite,%
\thanks{Also: Laboratorio de Astrof\'{ \i}sica Espacial 
y F\'{ \i}sica Fundamental,
Apartado 50727, 28080 Madrid} 
J. Margalef-Roig and S. Miret-Art\'es
\\ 
Instituto de Matem\'{a}ticas y F\'{ \i}sica Fundamental \\
Serrano 123, 28006 Madrid, Spain.}  

%
\maketitle

\begin{abstract}
In this work we classify the singularities obtained from the Gibbs potential 
of a lattice gas model 
with three components, two order parameters and five control parameters 
applying the general theorems provided by Catastrophe Theory. 
In particular, we clearly establish the
existence of Landau potentials in two variables or, in other words, corank 2
canonical forms that are associated to the {\it hyperbolic umbilic}, $D_{+4}$,
its dual the {\it elliptic umbilic}, $D_{-4}$, and the {\it parabolic 
umbilic}, $D_5$, catastrophes. The transversality of the potential 
with two order parameters is explicitely shown for each case. 
Thus we complete the Catastrophe Theory analysis of the three-component 
lattice model, initiated in a previous paper \cite{gaite2}.  
\end{abstract}
%
%



\global\parskip 4pt

\section{Introduction}
\label{Sec1}

The importance of phase transitions with several order parameters 
is very well known in
different branches of physics \cite{landau}.  A great amount of 
theoretical work has
been done in order to understand and construct accurate phase diagrams. 
As is well known, two different approaches are usually employed, one more
phenomenological by introducing Landau polynomial potentials which try to
describe experimental singular behaviors and the second one 
applying Catastrophe Theory (CT) (also known as Singularity Theory 
\cite{thom,poston,gilmore,okada}) and thus 
adopting a more methodological point of 
view. Even in the second case there are several ways to deal with phase
transitions and diagrams. Most works adopting the second point of view 
start with the canonical unfoldings as given and base their treatments on the
effect of perturbations leading to preserving or not the internal symmetry of
the system considered or, in other words, they focus their analysis in the
symmetry-breaking character of some phase transitions. 
The procedure we adopt here is
different since we begin with a thermodynamic potential (for example,
the Gibbs potential), assuming a mean field approach, and we apply
an algorithm or program according to the general mathematical theorems
established by CT, in order to extract all the topological information of the
original thermodynamics potential. In a recent paper we have applied this
CT program to a three component model phase diagram and we have found 
for a one order parameter potential the highest singularity with codimension 
five to be the {\it wigwam} or $A_6$ catastrophe \cite{gaite2}. 

Our CT program could be briefly stated as follows.
Let $H(x,\pi)$ be a real function with state variables $x_1,\dots,x_n$ ($x 
\in \IR^n$) and control parameters $\pi_1,\dots,\pi_r$ ($\pi
 \in \IR^r$); that is, $H : \IR^{n+r} \rightarrow \IR $. 
Then we must take the subsequent steps:

\begin{enumerate}

\item We pick ($x_0,\pi_0$) such that $x_0$ is a {\em degenerate critical
point} of $H(x,\pi_0)$ and we consider the unfolding
$h(x,\pi)=H(x+x_0,\pi+\pi_0)-H(x_0,\pi_0)$ and $g(x)=h(x,0)$  to
translate $x_0$ to the origin of coordinates. 

\item We calculate the determinacy and codimension of $g$ from the $k$-jet of 
$g$. Of course, if $g$ is $k$-determinate then $g \sim j^k(g)$, that is, the
function $g$ is equal to $j^k(g)$ up to a change of coordinates. In other
words, they are {\it equivalent} and have  qualitatively the same properties; 
therefore, both have the same codimension, cod$(g)=$ cod$(j^k(g))$.

\item We study the $k$-transversality of $h$. If this function is
$k$-tranversal we can affirm that $h$ and the canonical form of the unfolding
of $g$ are isomorphic. Then we can replace the original $H$ function by this 
canonical unfolding. If not, we can state that the $H$ function is not 
susceptible to be studied by CT.\\
\end{enumerate}

CT has usually not been applied in a rigorous way using all these concepts and
theorems needed for its correct implementation. This Catastrophe Program  
proposed here provides a very useful and systematic way to examine and 
classify  with not very much effort general behaviors of physical systems.
In particular, we emphasize the study of transversality of the
actual thermodynamical potentials  which guarantees that those simple forms 
(polynomial potentials or canonical forms)  represent  indeed 
up to a diffeomorphism the original thermodynamical potential. 
Following this CT program we do not need to invoke any convention (for
example, delay or Maxwell convention) in order to classify degenerate
or non-degenerate critical points on the state variables space.
Both conventions are not intrinsic to CT. Only when we deal with the time
evolution or when dynamical considerations about the physical system are 
considered, a given convention could be advisable. 
In particular, when the order of a phase
transition needs to be determined a convention is necessary 
because the transition occurs when
an appropriate separatrix in the control parameter space is crossed.

Here we will focus our attention on the
lattice gas model for a system with three components which simulates, in
particular, a binary fluid mixture. A wide literature has already been devoted
to it from different points of view \cite[and references therein]
{SST,griff2,Mei,griff1}, restricted to the case with one order 
parameter. In fact, very few studies with these methods can be found
for phase transitions with two order parameters \cite{schu2}.
This is rather surprising since the Landau potential for the three-state 
Potts model, which is a particularly important three-component model, 
has long been known to have two order parameters \cite{gaite1}.
We shall perform a complete CT study of the case with two order parameters, 
which is the maximum number for this model.

This work is organized as follows: In section 2 we describe the 
thermodynamical potential to be analyzed, 
give its physical interpretation and discuss general stability questions which
help connect usual concepts in Thermodynamics with those in Catastrophe Theory. 
In section 3 we apply the CT program to the potential previously introduced by
considering the singularities with corank equal to 2 and establish the
elementary catastrophes associated. In Section 4, we analyze in more detail the
Potts model as a particular case. 
The last section is devoted to a discussion
of the previous results and of the structure of the phase diagram entailed by 
them.

\section{Description of the Gibbs potential}

In the mean field theory, the Gibbs potential is a function of the
concentration of two of the three components and depends on three
thermodynamical parameters, which can be taken as the temperature and the
chemical potentials of the two components, and on three molecular parameters.
The phase diagram deduced from this function is an accurate description of the
system, except close to the (multi)critical points, where fluctuations become
important and alter significantly the mean field theory predictions. For this
reason, the Gibbs potential has been the basis for determining the
overall phase diagram \cite{griff2,Mei}. 

According to Ref. \cite{griff2} a phenomenological model for a ternary mixture
is obtained by assuming the Gibbs potential in the form 
\begin{equation}
\label{GIBBS}
{\bar G} = N [a' yz + b' xz + c' xy + R\,T\,(x \ln x + y \ln y + z \ln z)],
\end{equation}
where $N=N_x + N_y + N_z$ gives the number of total moles and $N_x$,
$N_y$ and $N_z$ the moles of each component. The variables $x,y,z$ 
are the mole fractions defined by $x=N_x/N$, $y=N_y/N$ and $z=N_z/N$; 
and hence we have the constraint 
\begin{equation}
\label{CONS}
x + y + z = 1 , \,\,\, \hbox{with} \, \, \, 0 < x, y, z < 1   .
\end{equation}  
This model can be derived from the mean field theory of a lattice model 
Hamiltonian with variables taking three different states, representing the 
molecules of the three components \cite{KM}. Then $a',b'$ and $c'$ represent 
molecular interaction parameters. The two order parameters to be considered 
in the sequel can be roughly thought of as the two independent differences 
between concentrations allowed by the constraint (\ref{CONS}).

Several systems of interest are described by this Gibbs potential, Eq.\ 
(\ref{GIBBS}): 
A ternary mixture at constant volumen, for example, a mixture of metals; 
a spin lattice where the molecules have spin one;   
a binary fluid mixture, where one of the three states represents a vacancy
instead of a new molecule and the corresponding concentration is associated to 
a variable total volumen. In the last case, the possible phases are
vapor, miscible liquid mixture and inmiscible liquid mixture. 
The convenient extensive variables are the specific volume $v$ and the relative 
concentration $\bar{x} = \frac{x}{x+y}$ of the two fluids and the intensive 
variables are the pressure and the chemical potential of one of the fluids. 
Moreover, the thermodynamical potential Eq.\ (\ref{GIBBS}) depends on $T,v$ and 
$\bar{x}$ and is therefore the Helmholtz potential $F(T,v,\bar{x}).$
This system is perhaps the most interesting
for applications, given the great amount of experimental data on 
binary fluid mixtures \cite{Row,McH-K}. However, the potential (\ref{GIBBS})
is not the most popular for fitting data; a related form which has 
similar dependence on the relative concentration of the two fluids but 
is of Van der Waals type for the volume is usually considered instead. We 
believe that this form, which is much more difficult to analyze, gives 
essentially the same qualitative behavior.

\section{Analysis of the Gibbs potential within the framework of
Catastrophe Theory}

Let us consider the reduced form of the Gibbs potential obtained from 
Eq.\ (\ref{GIBBS})  and dividing by $NRT$,
\begin{equation}
\label{RGIBBS}
G(x,y,z,a,b,c) = a\,yz + b\,xz +c\,xy + x \ln x + y \ln y + z \ln z
\end{equation}
where now the new parameters $a,b,c$ are defined with respect to 
the old ones $a',b',c'$ dividing them by $RT$. From the constraint 
Eq.(\ref{CONS}), we build a new function $H$ of two variables such that
\begin{eqnarray}
\label{HXY}
H(x,y,a,b,c) & = & a\, y (1-x-y) + b\, x (1-x-y) + c\,x y + x \ln x + y \ln y 
\nonumber \\
& + & (1-x-y) \ln (1-x-y)    .
\end{eqnarray}

The mean field theory prescription is then to minimize the {\it non-equilibrium
Gibbs potential} $H - \mu_x\,x - \mu_y\,y$, 
\begin{eqnarray}
\label{HHXY}
{\bar H}(x,y,a,b,c,\mu_x,\mu_y) & = & a\, y (1-x-y) + b\, x (1-x-y) + c\,x y 
+ x \ln x + y \ln y  \nonumber \\
& + & (1-x-y) \ln (1-x-y)  - \mu_x x - \mu_y y  ,
\end{eqnarray}
with respect to $x$ and $y$, 
where $\mu_x$ and $\mu_y$ are related to differences between the 
chemical potentials of the three components \cite{griff2}.   

CT will be applied to the ${\bar H}(x,y,a,b,c,\mu_x,\mu_y)$ function to 
classify the corank-2 singularities, at the generic point $(x_0,y_0)$
which moves on the triangle $x_0 > 0$, $y_0 > 0$ and $1 - x_0 - y_0 > 0$. 
CT conventionally uses the origin of coordinates as the point where 
singularities occur. Therefore, we shall 
translate the function ${\bar H}$ in order to have the singularities at the
origin. This translated function is written now as 
\begin{eqnarray}
\label{hxy}
h (x,y,a,b,c,\mu_x,\mu_y) & = &
{\bar H}(x+x_0,y+y_0,a+a_0,b+b_0,c+c_0,\mu_x + \mu_{x,0},\mu_y + \mu_{y,0}) 
\nonumber \\
& &{} - {\bar H}(x_0,y_0,a_0,b_0,c_0,\mu_{x,0},\mu_{y,0})  
\end{eqnarray}
and the germ of the unfolding $h$ is
\begin{equation}
\label{GERM}
g (x,y) = h(x,y,0,0,0,0,0)   .
\end{equation}
Now according to the CT program proposed in our previous paper \cite{gaite2} 
and mentioned in the Introduction, the following steps will be developed
for our problem: 

\begin{enumerate}

\item Find conditions for which the origin of coordinates is a degenerate
critical point of corank equal to 2 of $g$.

\item Show that $g$ is equivalent to $p$ (polynomial), shortly expressed as $g
\sim p$, which means that there is a diffeomorphism $\varphi$ such that 
$g = p\cdot \varphi$. The Determinacy and codimension of $g$ will also be evaluated.

\item Find a canonical unfolding ${\bar p}$ (polynomial) of $p$ by means 
of a basis of the vector space $\langle x,y\rangle / \Delta(p)$, 
which can be understood as the space of perturbations 
(see appendix B).

\item Find a canonical 5-unfolding ${\bar p}_1$ of $g$.

\item Establish the $k$-transversality of $h$ and if it holds,
then by using the main theorems about $k$-transversality, 
$h$ and ${\bar p}_1$ will be isomorphic.

\end{enumerate}

\subsection{Condition for a degenerate critical point at the origin.}

Now the first step is to write the conditions for which the origin of
coordinates (0,0) is a degenerate critical point with corank equal to 2
of $g$. This
can be done by equating all the first and second partial derivatives 
of $g$ to zero at the point (0,0) (hypothesis $\Sigma_0$)---thus 
the Hessian of the germ $g$ vanishes as well. 
This leads to five conditions among 
the variables and parameters $x_0,y_0,a_0,b_0,c_0,\mu_{x,0}$ and $\mu_{y,0}$. 
{}From that system of five equations the following relations can be extracted
\begin{eqnarray}
\label{TRI1}
a_0 & = & \frac{1}{2}\, (z_0^{-1} + y_0^{-1}) ,  \nonumber \\
b_0 & = & \frac{1}{2}\, (z_0^{-1} + x_0^{-1}) ,  \nonumber \\
c_0 & = & \frac{1}{2}\, (x_0^{-1} + y_0^{-1}) ,
\end{eqnarray}
and 
\begin{eqnarray}
\label{TRI2}
\mu_{x,0} & = & \frac{1}{2} (- y_0\, z_0^{-1} - x_0\, z_0^{-1} + 
z_0\, x_0^{-1} + y_0\, x_0^{-1}) + \ln x_0 - \ln z_0 ,  \nonumber \\
\mu_{y,0} & = & \frac{1}{2} (- y_0\, z_0^{-1} - x_0\, z_0^{-1} +
z_0\, y_0^{-1} + x_0\, y_0^{-1}) + \ln y_0 - \ln z_0,
\end{eqnarray}
with the definition $z_0 = 1 - x_0 - y_0 > 0$. Eqs.\ (\ref{TRI1})
give rise to a surface with parameters $x_0$ and $y_0$ fulfilling the
condition for a degenerate critical point with corank 2 at $(0,0)$ (Fig.\ 2). 

\subsection{Classification of the germ $g$.}

With the hypothesis $\Sigma_0$, the 3-jet of $g$ (Taylor expansion truncated
beyond terms of degree 3) around the point $(0,0)$ is
\begin{equation}
\label{JET}
j^3 (g) = \frac{1}{6}\, x^3\,(- x_0^{-2} + z_0^{-2}) + \frac{1}{2}\, 
x^2\, y\, z_0^{-2} + \frac{1}{2}\, x\, y^2\, z_0^{-2} + 
\frac{1}{6}\, y^3\,(- y_0^{-2} + z_0^{-2})    ,
\end{equation}
This 3-jet can be considered now as a homogenous 
polynomial of degree equal to 3 and can be rewritten as 
\begin{equation}
\label{CUB}
j^3(g) = a_1 x^3 + a_2 x^2 y + a_3 x y^2 + a_4 y^3   ,
\end{equation}
\noindent with $a_1 = (1/6)\,(- x_0^{-2} + z_0^{-2})$, 
$a_2=a_3=(1/2)\,z_0^{-2}$ and $a_4=(1/6)\,(- y_0^{-2} + z_0^{-2})$. 
We know from Singularity Theory that a general non-null homogeneous 
polynomial of degree 3 is equivalent by a linear 
transformation to one and only one of the following germs: 
$x^3 - x y^2$, $x^3 + x y^2$, $x^2 y$ and
$x^3$ \cite{ww,lander}. We must remark that 
these $x$ and $y$ are not to be identified with the initial physical 
variables though they are linearly related to them. 
The application of this lemma to Eq.\ (\ref{CUB}) 
to classify the 3-jet is given in  Appendix A and here only the final
conclusions will be summarized: 

\begin{enumerate}

\item  If $(1 - 2 x_0)\,(1 - 2 y_0) \, (1 - 2 z_0)  = 0$ then
$j^3 (g) \sim x^2 y$.

\item  If $(1 - 2 x_0)\,(1 - 2 y_0) \, (1 - 2 z_0) > 0$ then
$j^3 (g) \sim x^3 + x y^2$.

\item If $(1 - 2 x_0)\,(1 - 2 y_0) \, (1 - 2 z_0) < 0$ then
$j^3 (g) \sim x^3 - x y^2$.

\end{enumerate}

All of these three cases can be collected in a plot shown in Fig.\ 1. 
In Fig.\ 1 we 
display in the $x_0y_0z_0$ space the regions for the $x^3 \pm x y^2$ germs and
in Fig.\ 2 the separatrices among regions with different
codimensions in parameter space. The separatrix $x_0 = 1/2$, say, 
in conjuction with the equations defining the instability surface 
(\ref{TRI1}) gives rise to the plane $-a_0 + b_0 + c_0 = 2$, used in Fig.\ 2 
to find the separatrix in parameter space, which is a hyperbola branch.

\begin{figure}
\centering
\epsfxsize=7cm
\epsfbox{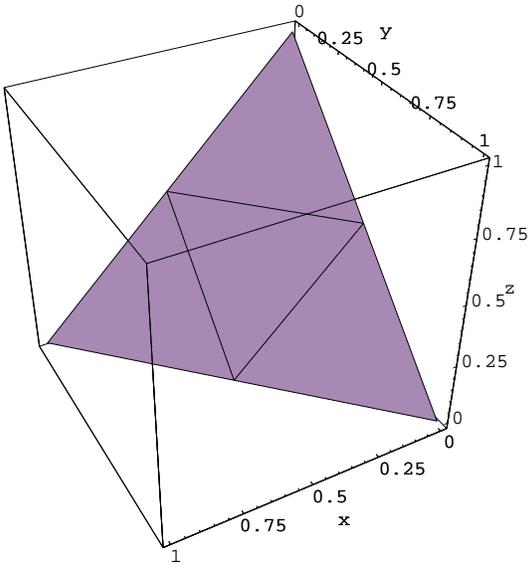}
\caption{Regions for $x^3 - x y^2$ potentials (inner triangle) and for 
$x^3 + x y^2$ potentials (3 outer triangles)}
\end{figure}
\begin{figure}
\epsfxsize=7cm
\epsfbox{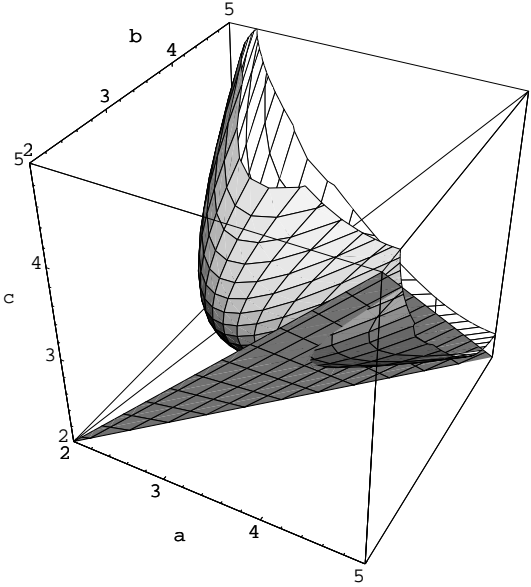}
\caption{Phase diagram showing the corank-2 instability surface. 
The intersection of the plane and the surface gives one of the three branches 
of the boundary between the $x^3 + x y^2$ and $x^3 - x y^2$ potentials. 
The diagonal corresponds to the 3-state Potts model. (See text)}
\end{figure}

The equivalence between the 3-jet and the canonical germs implies that their
codimensions are equal. Thus we have that
\begin{eqnarray}
\cod (j^3 (g)) & = & \cod (x^2 y) = \infty  , \nonumber \\ 
\cod (j^3 (g)) & = & \cod (x^3 - x y^2) = 3 , \nonumber \\
\cod (j^3 (g)) & = & \cod (x^3 + x y^2) = 3, 
\end{eqnarray}
and we observe that the behavior of the codimension of the 3-jet is
discontinuous according to Fig.\ 1. Indeed, when the point $(x_0,y_0)$ crosses 
the inner triangle the codimension of the
3-jet jumps to infinity. The codimension is only finite when what we could
call hypothesis $\Sigma_1$ is fulfilled, that is, 
\begin{equation}
\label{SIGMA1}
\Sigma_1 \equiv (1 - 2 x_0) (1 - 2 y_0) (1 - 2 z_0) \neq 0   .
\end{equation}
On the contrary, if $\Sigma_1 = 0$, corresponding to 
the sides of the inner triangle, the germ $g$ is not 3-determinate and 
we have to increase the order of
its jet by one more degree to explore 4-determinacy.   
We have that 
\begin{equation}
j^4 (g) = a_1 x^3 + a_2 x^2 y + a_3 x y^2 + a_4 y^3 +
b_1 x^4 + b_2 x^3 y + b_3 x^2 y^2 + b_4 x y^3 + b_5 y^4
\end{equation} 
where now the $b$-coefficients are found to be: $$b_1 = (x_0^{-3} +
z_0^{-3}) / 12,$$ $$b_2 = z_0^{-3} / 3,~~b_3 = z_0^{-3} / 2,~~b_4 = z_0^{-3} /
3,$$ $$b_5 = (y_0^{-3} + z_0^{-3}) / 12.$$
As before, in order to classify the 
4-jet of $g$ we invoke again a lemma of the Singularity Theory  
(see Appendix A). 
In all of the subcases examined the 4-jet is equivalent to the canonical
germ $x^2 y + y^4$, it is 4-determinate and its codimension is equal to 4.

\subsection{Determinacy and codimension of $g (x,y)$}

Once we have shown that the 3-jet is equivalent to the canonical germs 
$x^3 \pm x y^2$, and the 4-jet to $x^2 y + y^4$, 
it is clear that as these germs are 3- and 4-determinate, the 3-jet and 4-jet
will be also 3- and 4-determinate, respectively. 
Then $g \sim x^3 - x y^2$ (for $\Sigma_1 < 0$),  $g \sim x^3 + x y^2$ 
(for $\Sigma_1> 0$) both with $\cod(g) = 3$, and $ g \sim x^2 y + y^4$ 
(for $\Sigma_1 = 0$) but with $\cod (g) = 4$. These equivalences exist 
up to an unknown change of coordinates, so that here and in the following 
the variables $x$ and $y$ have no physical interpretation and may be 
regarded as dummy variables.

To see that $x^3 \pm x y^2$ are 3-determinate and
according to the main theorem about $k$-determinacy, it suffices to prove that 
\cite{ww,lander}
\begin{equation}
\label{DETER}
\langle x,y\rangle^{3+1} \subset \langle x,y\rangle^2\, \Delta(x^3 \pm x y^2) 
+ \langle x,y\rangle^{3+2}
\end{equation}
which is easy (for notation, see Appendix B). The same can be proven for
the 4-determinate case but  in Eq.\ (\ref{DETER}) the exponent $3$ should be
replaced by $4$.

\subsection{Canonical unfoldings of the germs $x^3 \pm x y^2$ and $x^2 y + y^4$}

It is well known from CT that $\{[x],[y],[x^2]\}$ is a basis for the quotient
vector spaces $\langle x,y\rangle / \Delta(x^3 - x y^2)$ and 
$\langle x,y\rangle / \Delta(x^3 + x y^2)$
and $\{[x],[y],[x^2],[y^2]\}$ for $\langle x,y\rangle / \Delta(x^2 y + y^4)$.
Moreover, the $k$-transversal (for all $k > 0$) canonical unfolding of 
the canonical form $x^3 - x y^2$ ({\it elliptic umbilic}, $D_{-4}$) 
and its dual $x^3 + x y^2$ ({\it hyperbolic umbilic}, $D_{+4}$) and $x^2 y +
y^4$ ({\it parabolic umbilic}, $D_5$)  are respectively
$$
x^3 - x y^2 + \lambda_1 x + \lambda_2 y + \lambda_3 x^2  ,
$$
$$
x^3 + x y^2 + \lambda_1 x + \lambda_2 y + \lambda_3 x^2  ,
$$ 
and 
$$
x^2 y + y^4 + \lambda_1 x + \lambda_2 y + \lambda_3 x^2  + \lambda_4 y^2.
$$ 

The corresponding bifurcation diagrams are well known in the Singularity Theory
\cite{LNM} and can be seen in any of the standard books on this theory 
\cite{poston,gilmore}. Equations governing such bifurcation diagrams are
$$
B_{D_{-4}} : 3 x^2 - y^2 + \lambda_1 + 2 \lambda_3 x = 0  , \, \,
	     - 2 x y + \lambda_2 = 0 , \,  \,
	     3 x^2 + x \lambda_3 + y^2 = 0       ,
$$
$$
B_{D_{+4}} : 3 x^2 + y^2 + \lambda_1 + 2 \lambda_3 x = 0  , \,  \,
	     2 x y + \lambda_2 = 0 , \,  \,
	     3 x^2 + x \lambda_3 - y^2 = 0      ,
$$
and 
$$
B_{D_5} : 2 x y + \lambda_1 + 2 \lambda_3 x = 0  , \, \,
	  x^2 + 4 y^3 + \lambda_2 + 2 \lambda_4 y = 0 , \, \,
        6 y^3 + y \lambda_4 + 6 \lambda_3 y^2 - x^2 + \lambda_3 \lambda_4 = 0,
$$
which are obtained by equating to zero the first
derivatives and Hessian of $g$ in each case.

\subsection{Canonical 5-unfolding of $g$}

Since $h$ has five parameters, the preceding unfoldings have to be extended
with two or one irrelevant parameters in order to apply the isomorphy theorem. 
Thus, for example, for the canonical unfoldings $x^3 \pm x y^2$, 
the two new unfoldings denoted by $\beta_1$ and $\beta_2$ can be written as 
\begin{equation}
\beta_1 (x,y,\lambda_1, \cdots, \lambda_5) = x^3 - x y^2 + \lambda_1 x +
\lambda_2 y + \lambda_3 x^2
\label{unfoldD-4}
\end{equation} 
and
\begin{equation}
\beta_2 (x,y,\lambda_1, \cdots, \lambda_5) = x^3 + x y^2 + \lambda_1 x +
\lambda_2 y + \lambda_3 x^2
\end{equation} 
and they are $k$-transversal unfoldings for all $k > 0$ of $x^3 - x
y^2$ and $x^3 + x y^2$, respectively. Now the new bifurcation sets are
$B_{\beta_1} = B_{D_{-4}} \times \IR^2$ and $B_{\beta_2} = B_{D_{+4}} \times 
\IR^2$.

In both cases, we can  affirm that there is a change of coordinates 
(a diffeomorphism) $\varphi_{(x_0,y_0)}$ 
such that $g = (x^3 - x y^2)\cdot\varphi_{(x_0,y_0)}$ and the same holds for 
$x^3 + x y^2$. Consequently, 
\begin{equation}
{\bar \beta_1} (x,y,\lambda_1, \cdots, \lambda_5) = s^3 - s t^2 + \lambda_1 s +
\lambda_2 t + \lambda_3 s^2
\end{equation}
and
\begin{equation}
{\bar \beta_2} (x,y,\lambda_1, \cdots, \lambda_5) = s^3 + s t^2 + \lambda_1 s +
\lambda_2 t + \lambda_3 s^2,
\end{equation}
where $(s,t) = \varphi_{(x_0,y_0)} (x,y)$, is a 3-tranversal unfolding
of $g$ with five parameters. Moreover, for the bifurcation sets we have that 
$B_{{\bar \beta_1}} = B_{\beta_1}$  and $B_{{\bar \beta_2}} = B_{\beta_2}$
A similar reasoning can be used for the canonical unfolding 
$x^2 y + y^4 + \lambda_1 x + \lambda_2 y + \lambda_3 x^2  + \lambda_4 y^2$ but
now we have only one irrelevant parameter.

Finally, these canonical unfoldings of $g$  need to
be related to the function $h$ or translated Gibbs potential. This is shown
explicitely in next subsection through the so-called transversality condition.

\subsection{$k$-transversality of the translated function $h$}

Now we return to the translated Gibbs potential $h$ to establish its 
relation with the unfolding of $g$ studied above. The condition for 
the existence of this relation is its transversality. Since we will 
be dealing with the physical function $h$ its arguments will be 
the original physical variables, to be distiguished from the dummy variables
used before. 
The vector space of the transversality, $V_h$, is defined by the first
partial derivatives of $h$ with respect to the five parameters according to 
\begin{eqnarray}
\label{TRANS}
V_h & = & \langle D_a h(x,y,0,0,0,0,0) - D_a h(0,0,0,0,0,0,0), \nonumber \\ 
& D_b & h(x,y,0,0,0,0,0) - D_b h(0,0,0,0,0,0,0),  \nonumber \\
& D_c & h(x,y,0,0,0,0,0) - D_c h(0,0,0,0,0,0,0),  \nonumber \\ 
& D_{\mu_x} & h(x,y,0,0,0,0,0) - D_{\mu_x} h(0,0,0,0,0,0,0), \nonumber \\ 
& D_{\mu_y} & h(x,y,0,0,0,0,0) - D_{\mu_y} h(0,0,0,0,0,0,0) \rangle_{\IR} 
\end{eqnarray}  
where $\langle \cdots\rangle_{\IR}$ means all the linear combinations with real
coefficients.

Let us analyze the transversality of the unfolding $h (x,y,a,b,c,\mu_x,\mu_y)$.
As was mentioned in the Introduction, this study is carried out 
in order to show that the function $h$ and the canonical unfolding of the 
germ $g$ are isomorphic. The vector space of
the transversality, $V_h$, is defined by Eq.(\ref{TRANS}) and reads in our case
\begin{equation}
\label{VH}
V_h = \langle - y x - y^2 + y z_0 - y_0 x - y_0 y, - x^2 - x y + x z_0 - x_0
x - x_0 y, x y + x y_0 + x_0 y, -x, -y\rangle_{\IR}   .
\end{equation}

Our next step is to prove that $h$ is 3- or 4-transversal, according to each
case. For this goal, we invoke 
one of the main theorems on $k$-transversality which states that $h$ will 
be $k$-transversal when the following algebraic condition is met
\cite{ww,lander}
\begin{equation}
\label{TRANS1}
\langle x,y\rangle = \Delta (g) + V_h + \langle x,y\rangle^{k + 1} ,
\end{equation}
which is fulfilled in our case for $k = 3$ or $4$. This requirement is
proved in Appendix C with the hypothesis $\Sigma_0$ and $\Sigma_1 \neq 0$ or 
$\Sigma_1 = 0$, respectively. We
finally conclude that $h$ is a 3- or 4-transversal unfolding of $g$ with five
parameters. Moreover, by using the main theorem on $k$-transversality, and for
the 3-transversal case, ${\bar \beta_1}$ and 
${\bar \beta_2}$ are isomorphic to 
$h$, that is, there  are three diffeomorphisms and a perturbation of 
parameters depending on $(x_0,y_0)$ for {\it each case} such that 
\begin{eqnarray}
\label{HXYXY}
h (x,y,a,b,c,\mu_x,\mu_y) & = &
{\bar H}(x+x_0,y+y_0,a+a_0,b+b_0,c+c_0,\mu_x + \mu_{x,0},\mu_y + \mu_{y,0}) 
\nonumber \\
&&{}- {\bar H}(x_0,y_0,a_0,b_0,c_0,\mu_{x,0},\mu_{y,0})  \nonumber \\
& = & s^3 \pm s t^2 + \lambda_1 s + \lambda_2 t + \lambda_3 s^2 +
\varepsilon_{(x_0,y_0)} (a,b,c,\mu_x,\mu_y), 
\end{eqnarray}
where the diffeomorphisms take the following expressions:
\begin{eqnarray}
\psi_{(x_0,y_0)} (x,y,a,b,c,\mu_x,\mu_y) & = &
(u,v,\lambda_1,\ldots,\lambda_5), \nonumber \\
\varphi_{(x_0,y_0)} (u,v) & = & (s,t),  \nonumber \\
\eta_{(x_0,y_0)} (a,b,c,\mu_x,\mu_y) & = & (\lambda_1,\ldots,\lambda_5), 
\end{eqnarray}
with
\begin{equation}
\psi_{(x_0,y_0)} (x,y,0,0,0,0,0) =  (x,y,0,0,0,0,0).
\end{equation}
All of these diffeomorphisms preserve the origin of coordinates.
Moreover, concerning the bifurcations, we have that
$\eta_{(x_0,y_0)} (B_{h}) = B_{D_{-4}} \times \IR^2$ and  
$\eta_{(x_0,y_0)} (B_{h}) = B_{D_{+4}} \times \IR^2$.
Similar expressions can be written for the 4-transversal case.

\section{The three-state Potts model}

The natural (and oldest) generalization of the 
Ising model consists of taking a site variable which can take three 
equivalent states 
instead of two, constituting the three-state Potts model. (Similarly, one 
can define the $q$-state Potts model.) It has complete permutation 
symmetry among the 
three states, yielding a Gibbs potential corresponding to 
(\ref{RGIBBS}) with $a=b=c$, Fig.\ 2, leaving as the only parameter the 
temperature $T$. The corank-2 instability occurs for $T_c=1/9=0.1111$ and the 
correponding potential is the $D_{-4}$ germ. The 3-parameter unfolding 
(\ref{unfoldD-4}) contains the possible perturbations of temperature or 
chemical potentials. Since in the three-state Potts model only the 
temperature perturbation is allowed we must have a one-parameter unfolding. 
Its form is best deduced by symmetry arguments \cite{gaite1}. 
We can substitute the germ by the symmetric form 
\begin{equation}
\label{3Pgerm}
z^3 + {\bar z}^3 = x^3 - 3\,x\,y^2,
\end{equation}
where we have introduced the complex variable $z = x+i\,y$. The permutation 
symmetry is obviously generated by the discrete rotations $z \rightarrow 
e^{i\,\frac{\pi}{3}}\,z$ and the reflection $z \rightarrow {\bar z}$.
The temperature perturbation preserves the symmetry and belongs to the 
vector space $\langle x,y\rangle / \Delta (j^3(g))$. 
Hence, it must be $x^2+y^2 = z\,{\bar z}$ (see Appendix B).
The corresponding unfolding is
\begin{equation}
\label{3Punfold}
z^3 + {\bar z}^3 + \lambda\,z\,{\bar z} = x^3 - 3\,x\,y^2 +\lambda\,(x^2+y^2),
\end{equation}
where $\lambda$ is a monotonic function of the temperature, at least, in 
a neighbourhood of $T_c$.

Now there arises the problem that the unfolding (\ref{3Punfold}), being a section 
of the complete unfolding of $D_{-4}$, contains potentials with one minimum 
and three saddle points if $\lambda>0$ or with one maximum 
and three saddle points if $\lambda<0$. These potentials are not bounded 
below nor have the three minima to be expected in this model. We must 
recall here the local character of CT and the discussion in \cite{gaite2}. 
There we remarked that the global character of these potentials can only 
be established by a numerical study over the entire range of the variables $x$ 
and $y$. Thus one finds that they indeed have three minima , far from 
the point $(0,0)$ and distributed symmetrically, if we keep the temperature 
in a neighbourhood of $T_c$. For a large value of $T$ only the minimum 
at $(0,0)$ survives. This is physically sensible, for at high temperature 
only the symmetric disordered phase can remain. The temperature $T_c$ 
precisely signals the point at which the symmetric disordered phase 
becomes unstable and disappears. Potentials for various situations are 
plotted in Fig.\ 3.

\begin{figure}
\centering
\epsfxsize=16cm
\epsfbox{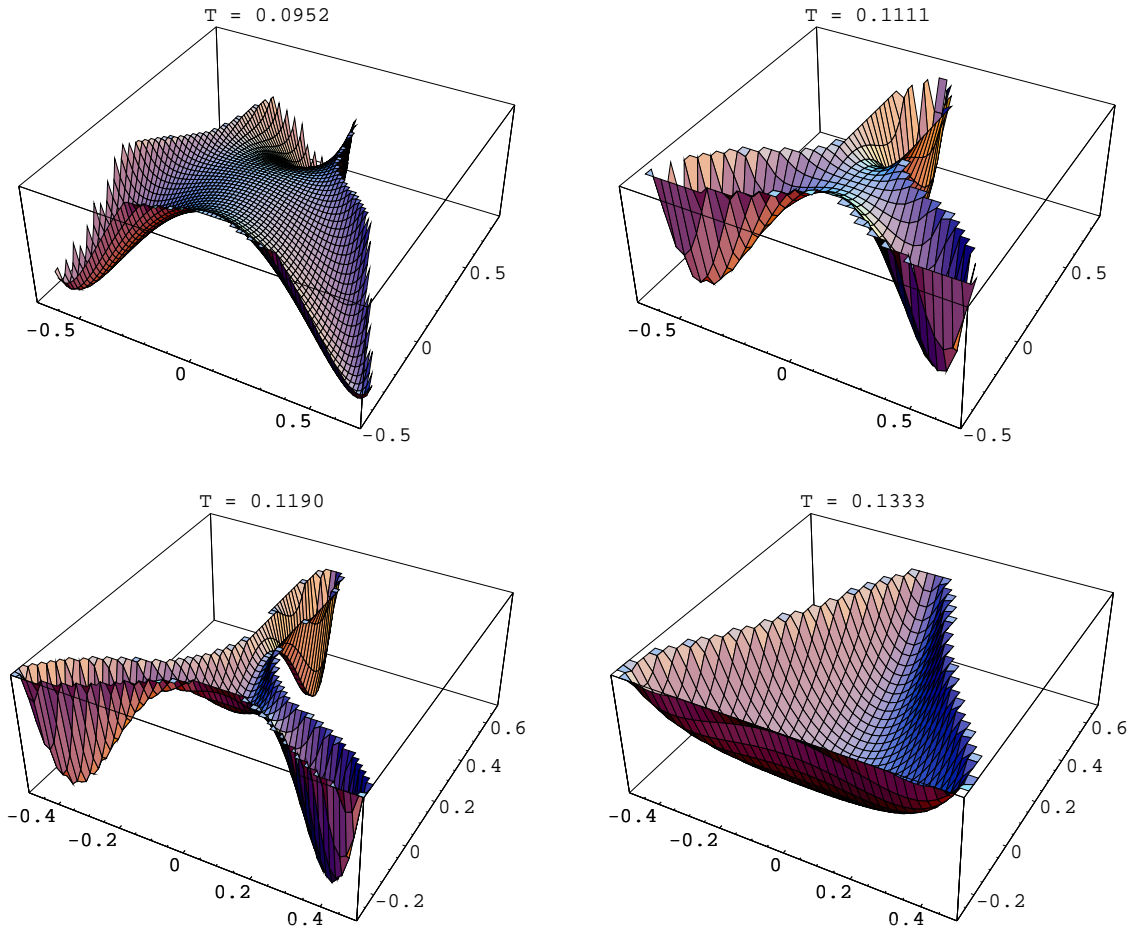}
\caption{Potentials for the 3-state Potts model at various 
temperatures.}
\end{figure}

Note that for $T>T_c$ the three minima converge to $(0,0)$, eventually 
merging there. Since their depth decreases as the depth of the 
minimum at $(0,0)$ increases, there must be a value of $T$ such that 
they are equal. If we adopt the Maxwell convention then we have 
a first order phase transition. This phase transition is known to 
occur in this model in space dimension $d=4$ and almost certainly for 
$d \geq 3$ \cite{Wu}. 
In $d=2$ the phase transition is definitely of second order \cite{gaite1}, 
which means that the mean-field theory potential with which we started 
is insufficient in low dimensions to describe accurately the real character 
of the possible phase transitions.

Landau potentials for the 3-state Potts model that are bounded below have 
appeared in the literature \cite{Golner,Zia-Wall,gaite1}. They are commonly 
derived by renormalization group (RG) arguments. In particular, it has been 
shown that for $d > 3$ the only RG relevant term further from the cubic is 
the quartic symmetric term $(x^2+y^2)^2 = (z\,{\bar z})^2$. It is even 
possible to prove with the powerful methods of $2d$ conformal field theory 
that this term, in addition to the cubic term, constitute a well defined 
Landau potential \cite{gaite1}. 
So the potential $c\,(z^3 + {\bar z}^3) +(z\,{\bar z})^2$ seems to be 
adequate for this model. In fact, due to the addition of the quartic term, 
it exhibits the same behavior as the total non-polynomial potential  
displayed in Fig.\ 3. In any event, we must remark that the quartic term does 
not belong to the germ, as calculated before. However, it could well arise 
from a more refined Gibbs potential, namely, of the type obtained with 
cluster variation methods \cite{Burley}, especially if symmetry 
arguments are invoked like in Ref.\ \cite{gaite1}.

\section{Conclusion}

We have analyzed the possible modes of instability and further 
singularities of the mean-field theory Gibbs potential for a three-component 
model in the case that two order parameters are needed for their 
description. Since the case of one order parameter has been already studied in 
Ref.\ \cite{gaite2}, the study of this Gibbs potential is now complete. 
We have found that three new catastrophes take place, namely, the 
{\it hyperbolic umbilic}, $D_{+4}$, its dual the {\it elliptic umbilic}, 
$D_{-4}$, and the {\it parabolic umbilic}, $D_5$, catastrophes. 
The {\it parabolic umbilic} has codimension 4, which is the highest we can reach 
in the case of two control parameters, unlike the case of one control 
parameter studied in \cite{gaite2}, where codimension five was reached.
The germ and 
unfolding of the {\it elliptic umbilic} catastrophe are precisely (isomorphic 
to) those of the 3-state Potts model, which is the model with the highest 
symmetry, and generally show three and even four phase coexistence. 
The total potential for the models belonging to the 
{\it hyperbolic umbilic} class do not have more than one local minimum and 
therefore they cannot give rise to phase coexistence. 
The phase structure for the {\it parabolic umbilic} is fairly complicated
and can be seen in any of the standard books about CT 
\cite{poston,gilmore,LNM}. 
Finally, it is remarkable that with two order parameters 
the highest codimension 5, which should give rise to even more complex  
singularities, is not reached.

We must say a few words about natural generalizations. Instead of limiting 
ourselves to three components we could well envisage the case of $q$ 
components. This would include most models ever considered in the physics 
of phase transitions. The most symmetric case is the $q$-state Potts model. 
An essential advantage of increasing $q$ is that mean-field theory becomes 
more accurate---in fact becoming exact for space dimension $d\geq 2$ if 
$q\geq 4$ \cite{Wu}.
A Landau potential for the $q$-state Potts model with $q-1$ 
order parameters is already known \cite{Zia-Wall}. Presumably, 
the rigorous study of the highest corank instabilities of the $q$ 
component Gibbs potential would produce a germ equal to the cubic term 
of that Landau potential, as for the 3-state Potts model. The 
corresponding singularities exhibit the interesting feature of 
having {\em modular parameters} \cite{ArnoldCS} already for $q=4$. 
There would certainly be a flock of other singularities with the same corank 
and the same or higher codimension, and with lower corank. We leave to 
the entrepreneurial reader the exploration of this endless world of 
mathematical entities and their physical application.    

\section{Acknowledgements}

This work has been supported by DGICYT-Spain with Grants PB96-0887,
PB96-0651-C03-01 and PB95-0071. 

\newpage

\appendix

\section{}

The application of the lemma of CT to Eq.(\ref{CUB}) to classify the 3-jet can
be carried out in a systematic way considering the following subcases: \\

\noindent ($A$) if $a_4 \neq 0$ (that is, $2 y_0 \neq 1 - x_0$), then 
the auxiliary cubic equation of Eq.(\ref{CUB}), obtained by dehomogenizing 
the polynomial, is given by
\begin{equation}
\label{POL3}
0 = \frac{a_1}{a_4} + \frac{a_2}{a_4} t + \frac{a_3}{a_4} t^2 + t^3.
\end{equation}
Its discriminant is defined as 
$\Delta_1^2=(t_1 - t_2)^2 \, (t_2 - t_3)^2\,(t_3 -
t_1)^2$, where $t_1$, $t_2$ and $t_3$ are the
roots of the cubic equation (\ref{POL3}). It can be expressed as 
\begin{eqnarray}
\label{DIS1}
\Delta_1^2 & = & - 4\, p^3\, r - 27\, r^2 + 18\, p\, q\, r - 4\, q^3 
+ p^2\, q^2 \nonumber \\
& = &\frac{1}{48\,{a_4}^4}\, x_0^{-4}\, y_0^{-4}\, z_0^{-4}\, 
(1 - 2 x_0)\,(1 - 2 y_0)\,(1 - 2 z_0)
\end{eqnarray}
with the following definitions: $p = - a_3/a_4$, $q= a_2/a_4$ and
$r=-a_1/a_4$. Notice the following subcases: \\

($A_1$) Eq.(\ref{POL3}) has three equal real roots ($\Delta_1^2 =
0$); this subcase is not possible because $y_0 > 0$, \\

($A_2$) Eq.(\ref{POL3}) has three real roots but two of them are
equal (that is, $\Delta_1^2 = 0$); in this subcase, $j^3 (g)$
and the monomial $x^2 y$ are equivalent or, in mathematical terms, 
$j^3 (g) \sim x^2 y$ (that is, the 3-jet and the 
monomial are equal up to a change of coordinates which essentially 
implies that $a_4 \neq 0$), \\

($A_3$) Eq.(\ref{POL3}) has three distinct real roots (that is, 
$\Delta_1^2 > 0$) and $j^3 (g) \sim x^3 - x y^2$, and  \\

($A_4$) Eq.(\ref{POL3}) has two conjugate complex roots (that is, 
$\Delta_1^2 < 0$) and then $j^3 (g) \sim x^3 + x y^2$, \\

\noindent ($B$) if $a_4 = 0$ and $a_1 \neq 0$ (that is, $2 y_0 = 1 - x_0$ and 
$2 x_0 \neq 1 - y_0$), then Eq.(\ref{POL3}) is replaced now by the auxiliary
equation
\begin{equation}
\label{POL33}
0 =  \frac{a_3}{a_1} t + \frac{a_2}{a_1} t^2 + t^3
\end{equation}
and its discriminant $\Delta_2^2$ is written as 
\begin{eqnarray}
\label{DIS2}
\Delta_2^2 & = & - 4\,q^3 + p^2 q^2 \nonumber \\
& = &  K(x_0,y_0) (2 -3 x_0 - 2 y_0) 
\end{eqnarray}
\noindent  with $K(x_0,y_0) > 0$.
The following redefinitions are now used: $p = - a_2/a_1$, 
$q= a_3/a_1$ ; $t_1$, $t_2$ and $t_3$ are again the new 
roots of the cubic equation, Eq.\ (\ref{POL33}). 
Thus we have again the following subcases\\

($B_1$) Eq.\ (\ref{POL33}) has three equal real roots. This is not possible 
since
$a_2 = a_3 \neq 0$, \\ 

($B_2$) Eq.\ (\ref{POL33}) has three real roots, two of them equal 
($\Delta_2^2 =
0$); this condition implies that $3 x_0 + 2 y_0 = 2$ and then $j^3 (g) \sim x^2
y$, \\

($B_3$) Eq.\ (\ref{POL33}) has three distinct real roots ($\Delta_2^2 > 0$), 
then $2 > 3 x_0 + 2 y_0$ and $j^3 (g) \sim x^3 - x y^2$, and \\

($B_4$)  Eq.\ (\ref{POL33}) has two conjugate complex roots ($\Delta_2^2 < 0$), 
then $2 < 3 x_0 + 2 y_0$ and $j^3 (g) \sim x^3 + x y^2$,  \\

\noindent ($C$) Finally we have the case where $a_4 = a_1 = 0$ (or $x_0 = y_0 =
1/3$) and we obtain that $j^3 (g) \sim x^3 - x y^2$. Due to the fact this case
is related to the well known Potts model, a more detailed  analysis of this
case will be addressed in Section IV. \\

For the 4-jet ($\Sigma_1 = 0$), the following subcases can be considered: \\

($D_1$) if $a_4 \neq 0 $ (that is, $2 y_0 \neq 1 - x_0$), then we take the 
auxiliary cubic equation 
\begin{equation}
\label{POL4}
0 = a_1 + a_2\, t + a_3\, t^2 + a_4\, t^3 .
\end{equation}
Eq.\ (\ref{POL4}) 
can have only three real roots of which two are equal. With the linear 
transformation
\begin{equation}
\label{PSI}
\psi^{-1} \equiv u = - a_4\, t_1\, x + a_4\, y , \, \, \, v = - t_2\, x + y
\end{equation}
with $t_1$ the single root and $t_2$ the double root of Eq.(\ref{POL4}), 
then 
\begin{equation}
j^3 (g)\,\psi (u,v) = u\,v^2.
\end{equation}
The value for the double root $t_2$ is easily
obtained from Eq.\ (\ref{POL4}) and its derivative with respect to the variable
$t$ since both equations are satisfied for $t_2$:
\begin{equation}
 t_2 = 
      {{\frac{  {a_2}\,{a_3} - 9\,{a_1}\,{a_4} }{2\,
	   \left(3\,{a_2}\,{a_4} -{{{a_3}}^2}
	      \right) }}} = 
	{{\frac{1 + 2\,x_0\,\left( -1 + y_0 \right)  - 2\,y_0}{2\,{{x_0}^2}}}}
= \left\{ \begin{array}{ll} -2\, y_0 & \mbox{if $x_0=1/2$}\\
			  \frac{-1}{2\,x_0}  & \mbox{if $y_0=1/2$}\\
			  -1 + {\frac{1}{2\,x_0}}& \mbox{if $z_0=1/2$}
	\end{array}. \right.
\end{equation}
One can further obtain $t_1$ and it coincides with $t_2$ only on the 
vertices of the triangle, where a concentration vanishes. Therefore, 
this situation, which would lead to a different singularity with 3-jet 
$u^3$, is actually outside the domain we consider.

We can write the 4-jet in the new variables as 
\begin{equation}
j^4 (g)\,\psi (u,v) = u v^2 + c_1 u^4 + c_2 u^3 v + c_3 u^2 v^2 + c_4 u v^3 +
c_5 v^4  ,                           \label{j4}
\end{equation}
where the new coefficients $c_i$ are linear combinations of the $b_i$. 
Only the sign of the non-zero $c_1$ coefficient is needed since, 
according to the mentioned lemma, the 4-jet is equivalent to 
\begin{equation}
j^4 (g) \sim x^2\, y + \hbox{sign}(c_1)\, y^4,
\end{equation}
provided that $c_1 \neq 0$.
This comes from the fact 
that all the 4-degree monomials of the 4-jet (\ref{j4}) belong to 
the Jacobian ideal of $u\,v^2$ except precisely $u^4$ and therefore 
can be removed by a diffeomorphism.
{}From the linear transformation Eq.\ (\ref{PSI}), we have that
\begin{equation}
c_1 = \gamma^4 \left[b_1 + b_2 t_2 + b_3 t_2^2 + b_4 t_2^3 + b_5 t_2^4\right]
\end{equation}
with $\gamma = 1 / [a_4 (t_2 - t_1)]$. Therefore the sign of $c_1$ is given by
the factor inside the bracket. 
Substituting for $t_2$, 
$c_1= \gamma^4 \,x_0^{-4}$ and is always positive, for any of
the three factors given by $\Sigma_1 = 0$, that is: $y_0 = 1/2$, $x_0 = 1/2$
or  $z_0 =1/2$. Finally,
\begin{equation}
\cod(j^4 (g)) = \cod(x^2 y + y^4) = 4
\end{equation}

($D_2$) if $a_4 = 0$, $a_1 \neq 0$ (that is, $2 y_0 = 1 - x_0$, $2 x_0 \neq 1 -
y_0$) and $\Sigma_1 = 0$, then only one point $(x_0,y_0)$ needs to be studied, 
$x_0 = 1/2$ and $y_0 = 1/4$.  The corresponding auxiliary cubic equation is now\begin{equation}
\label{POL44}
0 = t^3 + \frac{a_2}{a_1} t^2 + \frac{a_3}{a_1} t  .
\end{equation}
The roots of this last equation are: $t_1 = 0$ and $t_2 = - 2$. With the linear
transformation
\begin{equation}
\label{PSII}
\psi^{-1} \equiv  u = x + 2 y  , \, \, \,  v= 2 x
\end{equation} 
applied to the 4-jet we obtain that
\begin{equation}
j^4 (g)\,\psi(u,v) = v u^2 + d_1 u^4 + d_2 u^3 v + d_3 u^2 v^2 + d_4 u v^3 + d_5
v^4  .
\end{equation}
Here the sign of $d_5$ is positive and therefore the 4-jet is again
equivalent to the canonical germ $x^2 y + y^4$, its codimension being again
equal to 4.

\section{} 

In this Appendix we are going to present the mathematical concepts and 
notation widely introduced in Refs.\ \cite{ww,lander} and 
necessary to follow the main steps developed in Section III.

Let us consider real functions of class $\infty$ and defined in a neighbourhood
of $0 \in \IR^n$. We establish that two functions are equivalent if they
coincide in a neighbourhood of $0$. The classes we obtain are called 
{\em germs} of functions and the set of {\em germs} is denoted by $E(n)$. 
The operations $f+g$ and
$f\,g$ give to $E(n)$ the estructure of a ring and $M(n) = \{ f \in E(n) /
f(0) = 0 \}$ is a maximal ideal of this ring. Moreover, the operations 
$f+g$ and $\lambda \, f$ with $\lambda \in \IR$ give to $E(n)$ the 
structure of a real vector space of dimension $\infty$.  
The ideal $M(n)$ is generated by $x_1,\dots,x_n$, that is, 
$M(n) = \{ f_1 x_1 + \cdots + f_n x_n \;/\, f_1,\dots,
f_n \in E(n) \}$. In general, if $f_1,\dots,f_n \in E(n)$, we designate by
$\langle f_1,\dots,f_n \rangle $ the ideal generated by $f_1,\dots,f_n$,
that is, 
$$
\langle f_1,\dots,f_n \rangle = \{ f_1 g_1 + \cdots + f_n g_n \,\, /
\, g_1,\dots,g_n \in E(n) \}  .
$$
In particular, $M(n) = \langle x_1,\dots,x_n \rangle$.

It is possible to define powers of $M(n)$ as $M(n)^k$. It can be proven that
$M(n)^k$ is equal to the ideal of $E(n)$ generated by the monomials in $x_1,
\dots,x_n$ of degree $k$. In particular, for example, $\langle x,y \rangle ^2 = 
\langle x^2,xy,y^2 \rangle$ and
\begin{equation}
\langle x,y\rangle^2 = \{ f_1\, x^2 + f_2\,x y + f_3\,y^2\;/\,f_1, f_2, f_3 \in E(2) \}  .
\end{equation} 
We have also that 
$M(n)^{k+1} = \{f \in E(n)\;/\, D^if(0) = 0,~i \leq k \}$ where by $D^i$
we mean the derivative of degree $i$.

Now the main theorem about $k$-determinacy  establishes a sufficient condition
for a germ $g$ to be $k$-determinate. This condition reads 
\begin{equation}
\langle x,y\rangle^{k + 1} \subset  \langle x,y\rangle^2 . \Delta (g) + \langle x,y\rangle^{k+2}  , \label{kdeter}
\end{equation}
and, in our case, it will be applied for $k=3$ and $4$. 
The factor $\Delta (g)$ is known as the ideal of Jacobi and is
defined from the first partial derivatives of $g$ with respect to $x$ and $y$
($D_x g$ and $D_y g$, respectively). The calculation of this ideal is the
general starting point to establish the determinacy and 
codimension of $g$. Thus, for example, 
\begin{eqnarray}
\label{IJG}
\Delta (g) & = & \langle D_x g(x,y) , D_y g(x,y) \rangle  \nonumber \\
& = & \langle \frac{1}{2} x^2 M + x y z_0^{-2} + 
\frac{1}{2} y^2 z_0^{-2} + \frac{1}{3} x^3 (x_0^{-3} + z_0^{-3}) +
x^2 y z_0^{-3} + x y^2 z_0^{-3} + \frac{1}{3} y^3 z_0^{-3} +
r(x,y),  \nonumber \\ 
&&  \frac{1}{2} \,x^2 z_0^{-2} + x y\, z_0^{-2} + \frac{1}{2} y^2 N
+ \frac{1}{3} x^3 z_0^{-3} + x^2 y z_0^{-3}
+ x y^2 z_0^{-3} + \frac{1}{3} y^3 (y_0^{-3} + z_0^{-3}) + s(x,y)\rangle, 
\end{eqnarray}
with the definitions $M = - x_0^{-2} + z_0^{-2}$ and 
$N = - y_0^{-2} + z_0^{-2}$,
and with $r(x,y) \, , \, s(x,y) \in \langle x,y\rangle^4$ after Taylor 
expansions of $D_x g$ and $D_y g$ around the point $(0,0)$ have been 
performed, 
the first terms being ignored according to the  hypothesis $\Sigma_0$.
Analogously, we have that
\begin{equation}
\Delta (x^3 \pm x y^2) = \langle 3 x^2 \pm y^2, \pm 2 x y\rangle      .
\end{equation}

One is not always in the position to invoke a given lemma from CT to classify
the germ and establish its $k$-determinacy. In what follows we illustrate 
how to proceed in the general case using the condition above 
(\ref{kdeter}). Thus, 
\begin{eqnarray}
\langle x,y\rangle^2 \, \Delta (g) & + & \langle x,y\rangle^5  =  
\langle x,y\rangle^2 \, \langle \frac{1}{2} x^2 M + x y z_0^{-2} + 
\frac{1}{2} y^2 z_0^{-2},  \nonumber \\
& \frac{1}{2} & x^2 z_0^{-2} + x y z_0^{-2} + \frac{1}{2} y^2 N \rangle
 + \langle x,y\rangle^5  \nonumber \\ 
& = &  \langle x,y\rangle^5 + \langle \frac{1}{2} x^4 M + x^3 y z_0^{-2} + 
\frac{1}{2} x^2 y^2 z_0^{-2}, \nonumber \\
&& \frac{1}{2} \, x^4 z_0^{-2} + x^3 y z_0^{-2} + 
\frac{1}{2} x^2 y^2 N, \nonumber \\ 
&& \frac{1}{2} \, x^2 y^2 M
+ x y^3 z_0^{-2} + \frac{1}{2} y^4 z_0^{-2}, \frac{1}{2} x^2 y^2
z_0^{-2} + x y^3 z_0^{-2} + \frac{1}{2} y^4 N ,\nonumber \\
&& \frac{1}{2} \, x^3 y M + x^2 y^2 z_0^{-2} + 
\frac{1}{2} x y^3 z_0^{-2}, \nonumber \\
&& \frac{1}{2} \, x^3 y z_0^{-2} + x^2 y^2 z_0^{-2} + \frac{1}{2} x y^3 N\rangle, 
\end{eqnarray}
and therefore we have to prove that
\begin{eqnarray}
\mu_1 x^4 + \mu_2 x^3 y + \mu_3 x^2 y^2 + \mu_4 x y^3 + \mu_5 y^4 & = &
\nu_1 [\frac{1}{2} x^4 M + x^3 y\, z_0^{-2} + 
\frac{1}{2} x^2 y^2 z_0^{-2}] \nonumber \\
& + & \nu_2 [\frac{1}{2} x^4 z_0^{-2} + x^3 y\, z_0^{-2} + 
\frac{1}{2} x^2 y^2 N] \nonumber \\
& + & \nu_3 [ \frac{1}{2} x^2 y^2 M
+ x y^3 z_0^{-2} + \frac{1}{2} y^4 z_0^{-2}] \nonumber \\
& + & \nu_4 [ \frac{1}{2} x^2 y^2
z_0^{-2} + x y^3 z_0^{-2} + \frac{1}{2} y^4 N] \nonumber \\
& + & \nu_5 [ \frac{1}{2} x^3 y\, M + x^2 y^2 z_0^{-2} + 
\frac{1}{2} x y^3 z_0^{-2}] \nonumber \\
& + & \nu_6 [ \frac{1}{2} x^3 y\, z_0^{-2} + x^2 y^2 z_0^{-2} + \frac{1}{2} x y^3
N] 
\end{eqnarray}
and by equating coefficients of the same degree we obtain five
equations relating the $\mu$ coefficients with the $\nu$ coefficients. This
linear system of equations has solutions in $\nu_1, \ldots, \nu_6$ 
only if the hypothesis $\Sigma_1$ is fulfilled, Eq.(\ref{SIGMA1}).

Under these conditions, the germ and the 3-jet of the
germ are equivalent, $g \sim j^3 (g)$ and  $\cod(g) = \cod(j^3(g)) = 
\hbox{dim.\ vect.\ }\langle x,y\rangle / \Delta (j^3(g)) $, that is, the dimension of the quotient vector space
associated by the ideal of Jacobi of the 3-jet. Moreover, it can be also proved 
that $\{[x],[y],[x^2]\}$  is a basis of that quotient vector space.
Similarly, alternative basis can also be $\{[x],[y],[y^2]\}$ or 
$\{[x],[y],[x^2 + y^2]\}$.   

The same procedure can be applied to the 4-determinate case.

\section{}

As has been mentioned above we need to show that $h$ is 3-transversal
(similar calculations are needed in order to show that for the hypothesis
$\Sigma_1 = 0$, $h$ is 4-transversal). 
One of the main theorems about transversality establishes that this property 
is fulfilled when 
\begin{equation}
\langle x,y\rangle = \Delta (g) + V_h + \langle x,y\rangle^{3 + 1} .
\end{equation}
The ideal of Jacobi of $g$ is given by
Eq.\ (\ref{IJG}) and the vector space $V_h$ by Eq.(\ref{VH}). 

In Eq.(\ref{IJG}), the ideal of Jacobi of $g$ contains monomials of degree 
greater than 2. We are going to show that monomials of degree equal to 3 belong
to $\Delta(j^3(g))$. Thus we have that
\begin{eqnarray}
\lambda_1 x^3 & + & \lambda_2 x^2 y + \lambda_3 x y^2 + \lambda_4 y^3 
\nonumber \\
& = &  (a x + b y) ( \frac{1}{2} x^2 M + x y z_0^{-2} + \frac{1}{2} y^2
z_0^{-2}) + (c x + d y) (\frac{1}{2} x^2 z_0^{-2} + x y z_0^{-2} 
+ \frac{1}{2} y^2 N)
\end{eqnarray}
and the linear system of equations obtained from equating
coefficients of the same degree has solutions in $a, b, c$ and $d$  if 
$\Sigma_1 \neq 0$ .

On the other hand, a straightforward consequence of what we have shown above is
that
\begin{equation}
\Delta (g) + \langle x,y\rangle^4 = \Delta (j^3(g)) + \langle x,y\rangle^4    .
\end{equation}
So finally we can rewrite the transversality condition as 
\begin{eqnarray}
\label{CON3}
\langle x,y\rangle & = & \langle \frac{1}{2} x^2 M 
+ x y z_0^{-2} +  \frac{1}{2} y^2 z_0^{-2},  
\frac{1}{2}\, x^2 z_0^{-2} + x y z_0^{-2} + \frac{1}{2}\, y^2  N \rangle
+ \langle x,y\rangle^4  \nonumber \\
& + & \langle - y x - y^2 + y z_0 - y_0 x - y_0 y, - x^2 - x y + x z_0 - x_0
x - x_0 y, x y + x y_0 + x_0 y, -x, -y\rangle_{\IR}   .
\end{eqnarray}
In other words, we have to find a set of $\nu$-parameters fulfilling
\begin{eqnarray}
\label{CON4}
\mu_1 x +  \mu_2 y + \mu_3 x^2 + \mu_4 y^2 + \mu_5 x y &=&
\nu_1 ( - y x - y^2 + y z_0 - y_0 x - y_0 y) \nonumber \\ 
& + & \nu_2 ( - x^2 - x y + x z_0 - x_0 x - x_0 y) \nonumber \\
& + & \nu_3 (x y + x y_0 + x_0 y) \nonumber \\
& + & \nu_4 (-x)  \nonumber \\
& + & \nu_5 (-y)      \nonumber \\
& + & \nu_6 \,[ \frac{1}{2} x^2 M 
+ x y \alpha_0^{-2} +  \frac{1}{2} y^2 \alpha_0^{-2} ] \nonumber \\
& + & \nu_7 \,[\frac{1}{2} x^2 \alpha_0^{-2} + x y \alpha_0^{-2} + 
\frac{1}{2} y^2 N]  .
\end{eqnarray}
Again by equating coefficients of the same degree we obtain a set of
equations between  the family of known  $\mu$ and  unknown 
$\nu$ parameters. The corresponding  system of equations has a matrix of  
rank equal to 5 and the transversality condition is fulfilled.
 
For the 4-transversal case, the procedure is entirely similar. Only the
following observation needs to be made. Monomials of degree 1 and 2 are
included in the vector space of transversality and monomials of degree 5
and higher are considered in the term $\langle x,y\rangle^5$. 
So the equivalent to
Eq.\ (\ref{CON4}) has to take into account monomials of degree 3 and 4. This
leads to a system  of nine equations and ten unknown parameters. The rank of
that system is 9.

\end{document}